\documentclass[aps, prl, reprint]{revtex4-2}

\usepackage{bm, amsmath, amsfonts, mathtools, physics}
\usepackage[dvipdfmx]{color}
\usepackage{here}
\usepackage{cleveref}
\usepackage{etoolbox}

\crefname{figure}{FIG.}{FIGS.}
\Crefname{figure}{FIG.}{FIGS.}
\crefname{equation}{Eq.}{Eqs.}
\Crefname{equation}{Eq.}{Eqs.}

\usepackage{color} 
\definecolor{green}{rgb}{0., 0.66, 0.52}
\definecolor{indigo}{rgb}{0.26, 0.33, 0.42}
\definecolor{black}{rgb}{0., 0., 0.}

\begin{document}

  \title{
    Active Matter under Cyclic Stretch: Modeling Microtubule Alignment and Bundling
    }
  \author{Takumi Tagaki${}^{1}$}
    \email{tagaki-takumi@g.ecc.u-tokyo.ac.jp}

  \author{Seiya Nishikawa${}^{1}$}

  \author{Shuji Ishihara${}^{1, 2}$}
    \email{csishihara@g.ecc.u-tokyo.ac.jp}

  \affiliation{
  ${}^1$~Graduate School of Arts and Sciences, The University of Tokyo, Komaba 3-8-1, Meguro-ku, Tokyo 153-8902, Japan \\
  ${}^2$~Universal Biology Institute, The University of Tokyo, Komaba 3-8-1, Meguro-ku, Tokyo 153-8902, Japan
  }

\begin{abstract}  
We investigate the behavior of self-propelled particles under cyclic stretching, inspired by the characteristic pattern dynamics observed in microtubule (MT) motility assays subjected to uniaxial cyclic substrate stretching. 
We develop a self-propelled particle model that incorporates the elastic energy acting on the filaments due to substrate deformation, successfully reproducing the experimentally observed MT patterns. 
Additionally, the general framework of the model enables systematic exploration of collective responses to various substrate deformations, offering potential applications in the manipulation of MT patterns and other active matter systems.
\end{abstract}
\maketitle

\section{INTRODUCTION}
\begin{figure}[b]
  \includegraphics[keepaspectratio,scale=1.]{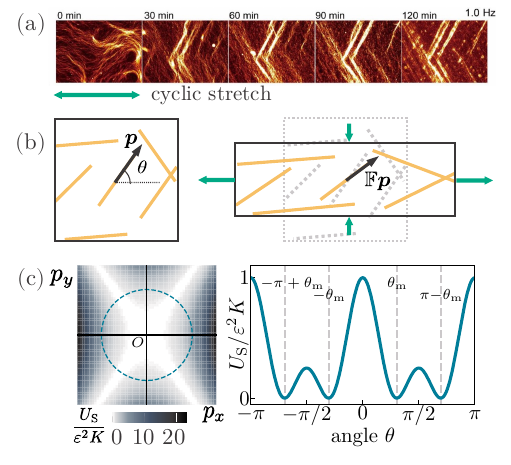}
  \caption{
    Effective energy for cyclic substrate deformation.
    (a) Fluorescence microscopy image of MTs on substrates subjected to uniaxial cyclic stretching.
    The green arrow shows the stretching axis.
    Adapted with permission from Inoue et al.~\cite{inoue2019}. 
    Copyright \copyright 2019 American Chemical Society.
    (b) Schematic illustration of the filament deformation by the substrate strain.
    (c) (left) Landscape of elastic energy $U_{\mathrm{S}}(\bm p)$ on ${\bm p} = (p_x,p_y)$. 
    The blue dashed line represents the $|\bm p| = 1$ circle.
    (right) $U_{\mathrm{S}}(\bm p)$ as a function of angle $\theta$ along the $|\bm p| = 1$ circle. 
    The gray dashed lines indicate $\theta = \pm \theta_{\rm m}$ and $\pi \pm \theta_{\rm m}$.
  }
  \label{fig1}
\end{figure}

  Microtubules play essential roles in various biological systems by exerting forces through interactions with motor proteins, kinesin and dynein~\cite{howard2001mechanics, shelley2016dynamics, hess2017nonequilibrium}. They drive cytoskeletal self-organization~\cite{ndlec1997self, surrey2001physical, karsenti2008self} and have inspired advances in synthetic microdevice engineering~\cite{sanchez2011cilia, sanchez2012spontaneous, sakuta2023self}.
  The MT motility assay, an \textit{in vitro} method of observing gliding MTs propelled by substrate-anchored motor proteins, has been used to investigate MT dynamics driven by motor proteins.
  The method has revealed collective dynamics of MT filaments such as vortices, turbulence, and bridging patterns~\cite{sumino2012vortex, torisawa2016spontaneous,ellis2018curvature,araki2021landscape, tsitkov2020kinesin,zhou2022durability}.
  Recent studies have shown that the MT dynamics can be effectively manipulated by mechanical perturbations, including regulation of attractive interaction among gliding MTs~\cite{inoue2015depletion} and confinement of MTs within regions with designed boundaries~\cite{araki2021landscape} or curved surfaces~\cite{hamant2019, ellis2018curvature}. 
  Inoue et al. reported the emergence of a unique MT pattern when uniaxial cyclic stretching was applied to the substrate, leading to MT alignment and propulsion at specific angles relative to the stretching axis, forming high-density bands (\cref{fig1}(a))~\cite{inoue2019}.
  While this experiment demonstrated how MTs collectively respond to external mechanical stimuli, the underlying mechanism behind the observed pattern formation remains unclear, raising two fundamental questions: (i) Why do specific angles emerge under cyclic stretch? (ii) Why do MTs form a banded pattern?

  Interestingly, similar alignment angles have been observed in cultured cells subjected to uniaxial cyclic stretching.
  To investigate cellular responses to mechanical forces generated by the heartbeat and respiratory rhythm, various uniaxial cyclic stretching experiments have been conducted using human umbilical vein endothelial cells \cite{takemasa1998}, aortic vascular smooth muscle cells \cite{zhu2011, hayakawa2000}, and fibroblasts \cite{livne2014, chatterjee2022, faust2011cyclic, wang2000comp}.
  These experiments demonstrated that the actin stress fibers within the cells, along with the elongated cell shapes, aligned in directions similar to those observed for MTs.
  This similarity between the alignment angles of MTs and those of cells or stress fibers under uniaxial stretching strongly suggests a common physical principle governing the responses to the mechanical deformation of the substrate.
  Although several theoretical studies have been conducted to explain the observed cellular responses under cyclic uniaxial stretching~\cite{livne2014, lucci2021, colombi2023}, there is no comprehensive theoretical framework integrating both the behaviors of cells and MTs.
  Moreover, previous studies have mainly focused on uniform substrate stretching, leaving the response of MTs and cells to more complex substrate deformations largely unexplored.

  To address the questions raised above, we developed a mathematical model to reproduce the observed MT pattern under the cyclic stretching (\cref{fig1}(a)) using the framework of active matter physics~\cite{marchetti2013hydrodynamics, chate2020dry, oByrne2022time,bar2020self}.
  We employ a self-propelled particle model, which has been successfully applied to explain the collective dynamics of MTs~\cite{sumino2012vortex, nagai2015memory}.
  First, we derive an effective energy representing the effect of externally applied cyclic stretching on the substrate, in reference to the theoretical approaches developed for cells~\cite{lucci2021, colombi2023}. 
  This energy determines the alignment angle approximated by $\theta_{\rm m} = \arctan(\sqrt2)$, known as ``magic angle'', an optimal fiber angle for enhancing material strength~\cite{horgan2022magicAngle, goriely2017}.
  Next, we introduce a self-propelled particle model that incorporates the external forces derived from the stretching. 
  This model reproduces the experimentally observed alignment angles and the banded aggregation patterns in MT systems, and provides further insights to the response of general active matter systems to the cyclic mechanical stimuli.

\begin{figure*}[t] 
  \includegraphics[keepaspectratio,scale=1.0]{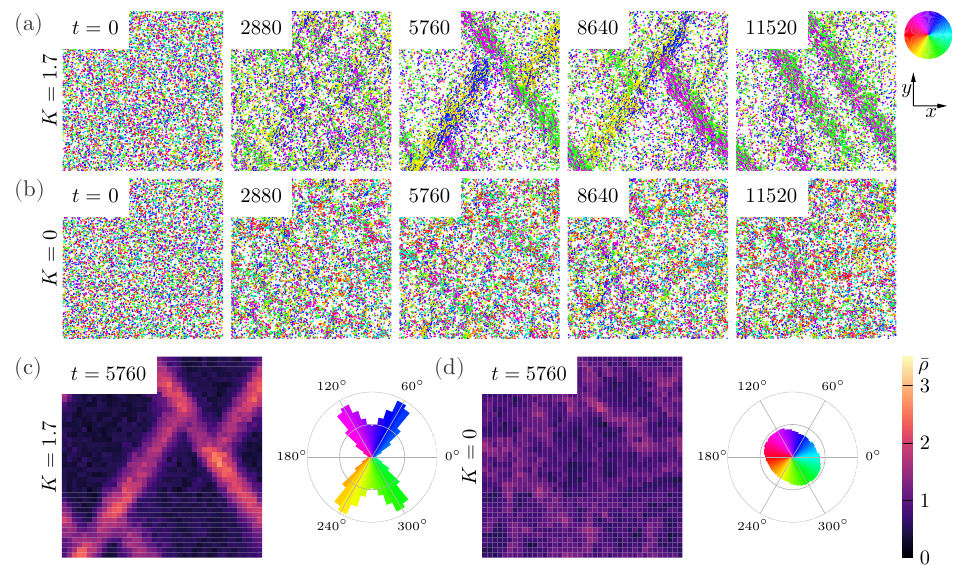}
  \caption{
    Numerical simulations of the self-propelled particle model.
    (a, b) Time evolution of the numerical simulations with and without external stretching, respectively. 
    $K$ is chosen as $K=1.7$ (a) and $K=0$ (b).
    Particle colors represent their directions $\theta_i$ depicted on the right. Stretch is applied along the $x$-axis.
    (c, d) Particle densities and angular distributions, corresponding to the data in (a) and (b) at $t=5,760$, respectively.
    Densities are displayed using the color scale indicated on the right.    
  }
  \label{fig2}
\end{figure*}

\section{METHOD}
\subsection{MTs alignment angle}
  Consider a MT filament bound to a substrate via motor proteins, as illustrated in \cref{fig1}(b). 
  When the substrate deforms, the point ${\bm X}_{\rm ref}$ in the reference configuration moves to ${\bm X} = {\bm X}_{\rm ref}+{\bm u}$, where ${\bm u}$ represents displacement.
  This substrate deformation induces deformation in the MTs involving rotation and either elongation or contraction; the MTs are subject to forces and energetically resist changes in their length.
  Let a polar vector ${\bm p} = (p_x, p_y)$ represent the length and direction of a MT filament.
  Under the condition that the substrate deformation is sufficiently faster than the MT motility, the MT filament state ${\bm p}$ in the reference configuration transforms to $\mathbb{F}{\bm p}$ owing to substrate deformation, where $\mathbb{F} \equiv \partial {\bm X}/\partial {\bm X}_{\rm ref}$ denotes the deformation gradient tensor~\cite{goriely2017}.
  The elastic energy to resist the length variation is given by
  \begin{align}
    \label{eq:Us_p}
    U_{\mathrm{S}} = \frac{K}{2} \left(  \left| \mathbb{F}\bm{p} \right|^2-\left| \bm{p} \right|^2 \right)^2 
    \simeq \frac{K}{2} \left\{\bm{p}^\mathsf{T} \left( \mathbb{E}^\mathsf{T} + \mathbb{E} \right) \bm{p} \right\}^2~,
  \end{align}
  where $K$ is the elastic modulus of the MTs related to resisting the length change, and superscript ${}^{\mathsf{T}}$ denotes the transpose of the vectors and tensors. 
  The energy $U_{\mathrm{S}}$ reaches the minimum value of zero when MT filament length remain unchanged by the substrate deformation, $\left| \mathbb{F}\bm{p} \right|=\left| \bm{p} \right|$.
  In the last expression of \cref{eq:Us_p}, we assumed that the deformation is small and retained only the leading order terms of the strain tensor, $\mathbb{E} \equiv \mathbb{F} - \mathbb{I} = \partial \bm u / \partial \bm X_\mathrm{ref}$, where $\mathbb{I}$ is the identity tensor.

  For uniaxial cyclic stretch along the $x$-axis, the strain tensor $\mathbb{E}$ is expressed as
  \begin{align}
    \mathbb{E} = 
      \varepsilon \sin (\omega t) \begin{pmatrix}
        1 & 0 \\
        0 & -\nu
    \end{pmatrix},
  \end{align}
  where $\varepsilon$ and  $\omega$ are the strain amplitude and frequency, respectively, and $\nu$ is the Poisson's ratio of the substrate.
  The energy $U_{\mathrm{S}}$ for the uniaxial stretch therefore becomes  
  \begin{align}
  \label{eq:Us_theta}
    U_{\mathrm{S}} 
      &= 2 \varepsilon^2 p^4 K \sin^2 (\omega t) \left( \cos^2 \theta - \nu \sin^2 \theta\right)^2 \nonumber\\
      &\simeq \varepsilon^2 p^4 K \left( \cos^2 \theta - \nu \sin^2 \theta\right)^2,
  \end{align}
  where we used parametrization $(p, \theta)$ defined by ${\bm p} = (p \cos \theta, p \sin \theta)^{\sf T}$. 
  The approximation in the last line was obtained by considering the time average, assuming that the substrate stretching occurs fast, as is the case in the experiment.
  The functional form of the energy is presented in \cref{fig1}(c).
  As the MT is rigid and the length $p$ hardly changes, the angle dependence of the energy is crucial and is shown in the right panel of \cref{fig1}(c).
  The energy reaches its minima at four angles, $\theta = \pm \theta_{\rm m}$ and $\pi\pm \theta_{\rm m}$, with $\theta_{\rm m} = \arctan (\sqrt{1/\nu})$.
  These angles can be easily interpreted.
  Vertically and horizontally aligned MTs are energetically unfavorable because they are subjected to contracting and tensile forces under uniaxial cyclic stretching, respectively. 
  MTs oriented at one of the four angles solely rotate without changing length.
  For an incompressible substrate for which Poisson's ratio is $\nu = 0.5$, $\theta_{\rm m} \simeq 54.74^\circ$, close to experimental observation~\cite{inoue2019}. 
  The energy $U_{\mathrm{S}} $ is the simplified form of that discussed for cells~\cite{lucci2021,colombi2023}.

\subsection{Self-propelled particle model}
  To address the formation of the MT bundling pattern observed in the experiments in \cite{inoue2019}, we developed a self-propelled particle model that incorporates alignment interactions and excluded volume effects~\cite{sumino2012vortex, nagai2015memory}.
  The model is composed of $N$ self-propelled particles in two-dimensional square region, where the position and the polar direction of $i$th particle at time $t$ are denoted by ${\bm r}_i(t)$ and $\theta_i(t)$, respectively ($i=1,\ldots, N$). 
  The dynamics of the system is described as follows:
  \begin{align}
    \label{eq:EqR}
    \frac{d\bm{r}_i}{dt} &= v_0 \bm{e}[\theta_i] - \frac{1}{\gamma} \sum_{i \neq j} \dfrac{\partial U_\mathrm{EV}(|\bm{r}_{ij}|)}{\partial \bm{r}_i},\\
     \label{eq:EqTheta}
    \frac{d\theta_i}{dt} &= \frac{\alpha}{\mathcal{N}_i} \sum_{\left| \bm{r}_j - \bm{r}_i \right| < R} \sin \left[ m (\theta_j - \theta_i) \right] + T_i + \xi_i.
  \end{align}
  Each particle is driven by a self-propelled force with constant speed $v_0$ along the direction dictated by the unit vector $\bm{e}[\theta] = (\cos \theta, \sin \theta)^\mathsf{T}$.
  The excluded volume effect among MTs is modeled using the Weeks-Chandler-Andersen potential~\cite{weeks1971perturbation}, defined by $U_\mathrm{EV}(r) = \lambda \left[ (d/r)^{12} - 2(d/r) ^{6}\right]$ for $r \leq d$ and $U_\mathrm{EV}(r) = -\lambda$ for $r < d$, where $\lambda$ and $d$ denote the interaction strength and the effective particle diameter, respectively. 
  The first term of \cref{eq:EqTheta} represents alignment interactions with the neighboring particles, where $\alpha$ is the interaction strength and $\mathcal{N}_i$ denotes the number of neighbors of the $i$th particle within distance $R$.
  For modeling MT dynamics, we mainly considered the nematic alignment interaction $m=2$, but also investigated polar alignment $m=1$ (see below).  
  $\xi_i$ represents white Gaussian noise with statistics of $\langle \xi_i(t)\rangle = 0$ and $\langle \xi_i(t) \xi_j(t')\rangle = \sigma^2\delta_{ij}\delta(t-t')$.
  $T_i$ in \cref{eq:EqTheta} represents the torque caused by the substrate deformation, derived from the stretching energy $U_{\mathrm S}$ discussed above,
  \begin{align}
    \label{eq:torque}
      T_i = -\left.\frac{1}{\eta}\frac{\partial U_\mathrm{S}}{\partial \theta}\right|_{p=1}~.
  \end{align}   
  
  We numerically solved the above stochastic equations (\cref{eq:EqR,eq:EqTheta}) in a $L \times L$ square region with periodic boundary conditions.
  The calculations were implemented using the Euler-Maruyama method, with a discretized time step $\Delta t = 0.001$. The parameters were set as follows unless otherwise stated: $N=10,000$, $L=100$, $\gamma = 1.0$, $\eta=1.0$, $v_0 = 0.1$, $\lambda=7.17 \times 10^{-6}$, $d=1.0$, $\alpha=0.953$, $R=0.5$, $m=2$, $K=1.7$, $\varepsilon=0.1$, and $\sigma=0.1$.
  $R$ was set smaller than $d$ because MTs are aligned when they collide with each other~\cite{hamant2019, frey2020continum, sumino2012vortex}.
  For the initial conditions, ${\bm r}_i$ and $\theta_i$ were randomly chosen from uniform distributions, respectively.
  
\begin{figure}[b] 
  \includegraphics[keepaspectratio,scale=1.06]{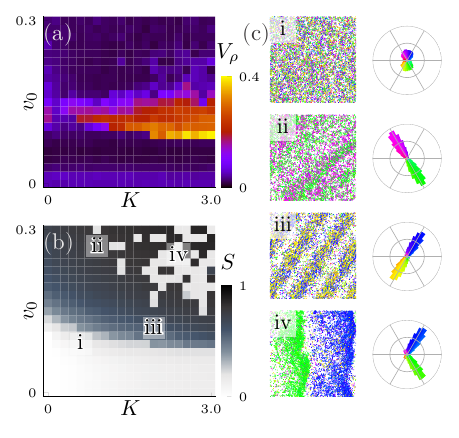}
  \caption{
    Phase diagram of the self-propelled particle model on the $K$-$v_0$ plane.
    (a) Density variance $V_{\rho}$ and (b) nematic order parameter $S$ shown by color maps indicated in the right.
    (c) Snapshots and angular distribution of $
    \theta_i$ for each pattern at the parameter (i)--(iv). Movies are available in Supplymentary Information MOV.~S2.
    (i) disorder, (ii) nematic, (iii) band, (iv) polar wave.
  }
  \label{fig3}
\end{figure}

\begin{figure}[t]
  \includegraphics[keepaspectratio,scale=1.0]{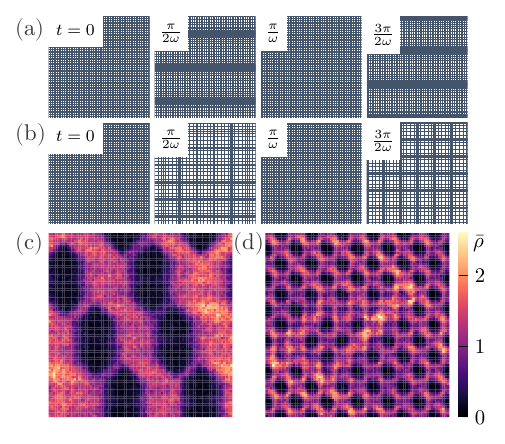}
  \caption{
    Pattern control by designed substrate deformation.
    (a, b) Illustrations of applied cyclic deformations \cref{eq:eq_heterodeformation}.
    At $t=0$ and $\pi/\omega$ the substrates are in the reference state ($\mathbb{E}^{\alpha}(t=0, \bm{X}_\mathrm{ref})=0$).
    The grid represents the displacement of the substrate over time.
    Note that the time-averaged effect of deformation was considered in the simulations.
    (c, d) Particle density maps obtained under the deformation shown in (a) at $t=54,000$ and 
    (b) at $t=108,000$, respectively.  
  }
  \label{fig4}
\end{figure}

\section{RESULTS OF NUMERICAL SIMULATIONS}

\subsection{Cyclic uniaxial stretching}  
  \cref{fig2}(a) exhibits typical time series from numerical simulations (Supplementary Information MOV.~S1).
  Starting from random initial conditions, individual particles align along one of the four angles $\pm \theta_{\rm m}$ and $\pi \pm \theta_{\rm m}$. 
  Subsequent collisions between particles result in the formation of nematic bands similar to the experimental observations, as depicted in the particle density map and angle distributions in \cref{fig2}(c).
  Here the density map $\overline{\rho}$ is calculated by dividing the system into a $40\times 40$ mesh and taking the time average of the particle number over $T=20$ for each mesh.
  The crossing points of these bands gradually shift over time, eventually forming a striped pattern where particles propel in both directions along the stripes.
  The formation of the bands contrasts with the case without cyclic stretching ($K=0$; \cref{fig2}(b) and MOV.~S1) where the particles remain uniformly distributed with respect to their position and propelling direction (\cref{fig2}(d)). 
  These results suggest that the application of cyclic stretch plays a crucial role in band formation.
 
  To understand the importance of cyclic stretching and self-propulsion, we studied the parameter dependence of the system dynamics by varying $K$ and $v_0$.
  We measured the particle density fluctuation (variance of the local density $\overline{\rho}$ over the system) $V_\rho$ and nematic order parameter $S = \left| (1/N) \sum_i e^{2i\theta_i} \right|$ for each parameter set, as shown in \cref{fig3}(a) and (b).
  \cref{fig3}(c) displays four snapshots, along with the angular distributions of the particles' propulsion directions, obtained from the simulations corresponding to the parameter sets indicated by Greek numerals in \cref{fig3}(b) (see also MOV.~S2).
  For small values of $v_0$, the local density remains uniform and alignment is not observed (i: disorder), indicating the significant role of motility in the onset of the MT banded pattern.
  As $v_0$ increases and exceeds a certain threshold, the particles begin to align.
  Under weak stretching (small $K$), the particles either align or anti-align with each other but remain uniformly distributed, as shown in (ii: nematic), where $S>0$ and $V_\rho \simeq 0$.
  As the strength of stretch $K$ increases, a region appears where $V_\rho$ and $S$ simultaneously take positive values, and a banded pattern is observed as represented in (iii: band).
  For higher values of $v_0$, the particles eventually separate into two populations, each directed along the same propulsion angle (iv: polar wave).
  The two populations do not collide with each other because the combination of the angles is $\theta_{\rm m}$ and $-\theta_{\rm m}$ (or $\pi-\theta_{\rm m}$ and $\pi+\theta_{\rm m}$ depending on the initial condition).
  Note that in this region the eventual state is highly dependent on the initial condition, and the system may evolve into a nematic state similar to (ii).
  These results indicate that both self-propulsion and stretching are important for the appearance of the banded pattern observed in the experiment of \cite{inoue2019}.
  
  We also studied the model with polar alignment, $m = 1$. 
  Our numerical simulations did not produce band formation across wide range of parameters (Supplementary Information FIGs.~S1 and S2).
  Earlier studies have suggested that polar alignment ($m=1$) leads to the formation of Vicsek waves, for which the band moves in a direction perpendicular to its elongation, whereas nematic interaction ($m=2$) is crucial for the formation of bands in which particles move parallel to the band length~\cite{frey2020continum}.
  Thus, we conclude that nematic interaction is required for the reproduction of the banded pattern observed in the MTs experiment.

  \subsection{Cyclic Stretching under General Strain Conditions}  
  So far, we have focused on self-propelled particles under cyclic uniaxial stretching using \cref{eq:Us_theta}; however, the stretching energy given in \cref{eq:Us_p} can accommodate arbitrary strain tensors $\mathbb{E}$.
  This formulation allows us to explore possible scenarios for controlling the patterns and dynamics of active matter systems, including MTs, which have not been experimentally investigated yet. 
  Here, we present two examples of numerical simulations that apply distinct cyclic deformation, $\mathbb{E}^{\alpha}(t, \bm{X}_\mathrm{ref}) = \varepsilon \sin (\omega t) \mathbb{E}^{\alpha}_0(\bm{X}_\mathrm{ref})$ ($\alpha = 1, 2$), with $\mathbb{E}^{\alpha}_0(\bm{x})$ chosen as
  \begin{align}
  \label{eq:eq_heterodeformation}
      \mathbb{E}^{1}_0
   \!=\!
   \begin{pmatrix}
      1 & 0 \\
      0 & \sin \frac{6\pi y}{L}
    \end{pmatrix}
    ~{\rm and}~
 \mathbb{E}^{2}_0
   \!=\!
   \begin{pmatrix}
       \sin \frac{12 \pi x}{L} & 0 \\
      0 & \sin \frac{12 \pi y}{L}
    \end{pmatrix},
  \end{align}  
  respectively.
  These deformations are not uniform in space, as illustrated in \cref{fig4}(a) and (b).
  We performed numerical simulations with $N=40,000$ and $L=200$ to keep the total particle density. 
  The numerical simulations resulted in patterns that were not straightforwardly predictable.
  The deformation $\mathbb{E}^{1}$ resulted in hexagonal and thick band patterns as shown by particle density map $\overline{\rho}$ in \cref{fig4}(c), whereas the deformation $\mathbb{E}^{2}$ led to oblique square lattice patterns, also depicted in \cref{fig4}(d).
  These examples provide testable predictions of the current model for future experiments, while offering a method to control the patterns of active matter through mechanical stimuli.
  
\section{DISCUSSION}
  In summary, we developed a mathematical model for self-propelled particles subjected to cyclic deformation by incorporating elastic energy to resist changes in length.
  For uniaxial cyclic stretching, our model successfully explained the alignment angle and reproduced the banded pattern observed in MT motility assays~\cite{inoue2019}.
  The elastic energy, which is reflection-symmetric with respect to the $x$- and $y$-axes, has minima at four specific angles. 
  This constrains the preferential propulsion directions of the particles, leading to instability in the uniform particle density distribution and the emergence of a banded pattern. 
  Our simulations revealed that an appropriate self-propulsion (intermediate values of $v_0>0$), stretching strength (large enough $K$), and nematic interaction ($m=2$) are important for the banded pattern formation.
  These findings should be further explored, not only through improving the modeling accuracy of actual experiments, such as employment of filamentous particles~\cite{bar2020self, vliegenthart2020filamentous},
  but also by developing more analytical approaches, such as continuum modeling.
  For instance, the effects of cyclic stretching can be integrated into existing continuum models for MTs~\cite{frey2020continum, peshkov2012nonlinear}, by which we could analytically determine the boundary between different dynamic phases in~\cref{fig3}.

  The present formulation of the elastic energy given in~\cref{eq:Us_p} is adaptable to a wide variety of substrate strains.
  While most experimental studies have focused on uniform deformation, particularly cyclic uniaxial stretching, MTs and cells \textit{in vivo} experience more complex substrate deformations.
  Our model is useful for investigating the response of MTs to such complex substrate deformations. Furthermore, the model offers a potential tool for manipulating MT patterns and dynamics, as demonstrated in~\cref{fig4}.
  Since our approach is based on a general theoretical framework for active matter physics and does not depend on the specific properties of MTs, it can be extended to and compared with other systems.
  In our model, the effect of cyclic substrate deformations is represented as an external potential acting on the angular variables of self-propelled particles.
  A similar modeling scheme was recently applied to bacteria under a magnetic field~\cite{beppu2024magnetically} and cells on substrates with micro-fabricated shallow ridges~\cite{zhao2024external}.
  Although the form and symmetry of the potential energy differ among these systems, exploring the underlying physical principles through modulation of angular variables is a promising theoretical strategy for effective and programmable control of active matter~\cite{PhysRevE.108.044601}. We hope this study stimulates future research in both theoretical and experimental directions.

\section*{ACKNOWLEDGMENTS}
  We acknowledge A. Kakugo, M. Tani, K. Mitsumoto and T. Namba for their fruitful discussions and kind supports. This study was supported by JSPS KAKENHI Grant Numbers 24H01931 (to S.I.) and JST CREST JPMJCR1923, Japan (to S.I.). 

\section*{Author Contributions}
  All authors devised the project. T.T. performed the simulations and analyses supervised by S.S. and S.I. T.T. and S.I. wrote the manuscript. 

\section*{Data Avairability}
  The code used for the numerical simulations is openly available at:
  \url{https://github.com/IshiharaLab/AMCyclicStretch}.





\bibliographystyle{apsrev4-2}
\bibliography{main}
  
\end{document}